\documentclass[pre,twocolumn,english,showpacs,aps,prb,a4paper,floatfix]{revtex4}
\usepackage[T1]{fontenc}
\usepackage{color}
\usepackage[latin1]{inputenc}
\usepackage{graphicx}
\usepackage{bm}
\usepackage{epsfig}
\usepackage{hyperref}

\begin{document}
\title{Domain walls in two-dimensional nematics confined in a small circular cavity}
\author{Daniel de las Heras$^1$}
\author{E. Velasco$^2$}
\address{$^1$ Theoretische Physik II, Physikalisches Institut,
Universit\"at Bayreuth, D-95440 Bayreuth, Germany\\
$^2$ Departamento de F\'{\i}sica Te\'orica de la Materia Condensada and Instituto de F\'{\i}sica de la Materia Condensada,
Facultad de Ciencias, Universidad Aut\'onoma de Madrid, 
E-28049 Madrid, Spain}

\begin{abstract}
Using Monte Carlo simulation, we study a fluid of two-dimensional hard rods inside a small circular cavity bounded by 
a hard wall, from the dilute regime to the high-density, layering regime. Both planar and homeotropic anchoring of the nematic director 
can be induced at the walls through a free-energy penalty. The circular geometry creates frustration in the nematic phase 
and a polar-symmetry configuration with a distorted director field plus 
two $+1/2$ disclinations is created. At higher densities, a quasi-uniform structure is observed with a (minimal) director 
distortion which is relaxed via the formation of orientational domain walls. This novel structure is not predicted by elasticity theory and
is similar to the step-like structures observed in three-dimensional hybrid slit pores. We speculate that the formation
of domain walls is a general mechanism to relax elastic stresses in conditions of strong surface anchoring and 
severe spatial confinement.
\end{abstract}

\maketitle

\section{Introduction}

In adsorbed nematics, surfaces often determine the favoured director orientation, which then propagates into the bulk
\cite{Jerome}. When a nematic is subject to orientations propagating from different surfaces, elasticity theory predicts 
a distorted director configuration, with an associated elastic free energy. Very often conflicting surface orientations 
frustrate the director in restricted geometries and disclinations are generated. Disclinations can be generated by curvature 
alone \cite{Lubensky}, for example by placing a two-dimensional (2D) nematic on a finite but unbounded surface such as a spherical surface 
\cite{Bowick,PhysRevLett.108.057801,Lowen}. The dimensionality of space is also important since it forces the possible topology and limits the 
type of defects that can form. 

Consider a 2D nematic made of rod-like particles inside a small (of a size a few times the particle length) 
circular cavity. When the density is increased from a dilute state, the bulk isotropic-nematic transition \cite{Cuesta,Bates,review,Lucio}
induces some kind of orientational ordering in the cavity. However, due to the surface, the nematic director is 
unable to adopt a defect-free uniform configuration, and a global topological charge $+1$ \cite{Nelson} arises in the cavity: 
either as a central $+1$ disclination, or as two diametrically opposed $+1/2$ disclinations (on the surface or at a distance
from it). Thus the fluid optimises the surface free energy at the cost of creating two disclinations and a distorted director 
field (which incurs an elastic free energy). This effect has been observed in Monte Carlo (MC) simulations \cite{Lowen}. For
weak surface anchoring there is an alternative scenario: the director may not follow the favoured surface orientation 
and a quasi-uniform, defect-free configuration with little elastic free energy may arise in the cavity. A uniform phase has 
been obtained in vibrated monolayers of granular rods \cite{Galanis}. Interesting phase diagrams result as the cavity radius 
and the fluid density are changed, as exemplified by recent density-functional calculations \cite{PhysRevE.79.061703,herasliqcrys,Chen}.

\begin{figure}[h]
\includegraphics[width=3.0in,angle=0]{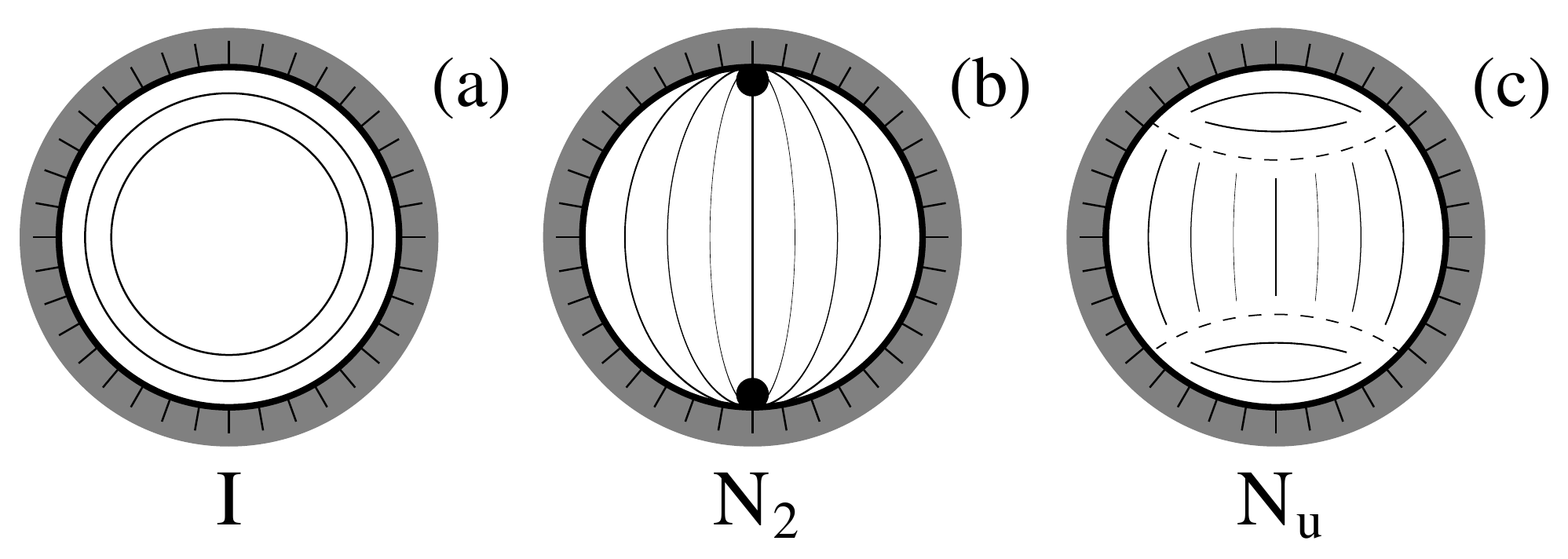}
\caption{Schematic of the phases obtained in the circular cavity as the density is increased in the case of a wall 
imposing planar surface alignment. Continuous thin line is the director field. (a) Low density: Isotropic phase (I) with a 
film of nematic fluid adsorbed on the surface. (b) Intermediate density: Polar phase (N$_2$) containing two disclinations 
(in the case depicted disclinations are at the wall). (c) High density: Quasi-uniform (N$_u$) phase with two domain walls 
(represented by dashed lines).}
\label{fig1}
\end{figure}

In this paper we use Monte Carlo simulation to analyse this system, using hard rectangles of length-to-width ratio $L/\sigma$ 
as a particle model and a circular cavity of radius $R$ that imposes hard or overlap forces on the particles. Both planar 
(tangential) and homeotropic (normal) anchorings are studied. In both cases we observe the formation of 1D domain
walls when the cavity radius is small. The scenario is schematically depicted 
in Fig. \ref{fig1} (for the sake of illustration, only the planar case is depicted): an isotropic phase, I, at low density 
(containing a thin nematic film in contact with the wall), Fig. \ref{fig1}(a), followed, at higher densities, by a nematic 
phase with two $+1/2$ disclinations called polar configuration, N$_2$, see Fig. \ref{fig1}(b). When the density 
increases further the polar configuration is no longer stable, and may transform into a structure with a quasi-uniform 
director configuration, phase N$_u$ in Fig. \ref{fig1}(c). Depending on the surface conditions, cavity radius and 
particle aspect ratio, the N$_u$ phase may be preempted by formation of layered (smectic-like) structures. 

Our study contains two novel features. (i) In contrast with previous works, which focus solely on the nematic phase
and the configuration of disclinations, the whole density range, from dilute to near close-packing, is explored here.
(ii) The peculiar structure of the N$_u$ phase: the incorrect surface alignment associated with a uniform director is here
observed to be relaxed by the formation of domain walls, i.e. one-dimensional interfaces across which the director rotates abruptly
by approximately $90^{\circ}$; this is represented by the dashed lines in Fig. \ref{fig1}(c). These structures
are similar to the step-like defects observed in three-dimensional nematics inside hybrid planar slit pores 
\cite{PM,Galabova,Sarlah,PhysRevE.79.011712,paulo,Noe} or near half-integer disclinations
in cylindrical pores \cite{Schopohl}, and may be a universal feature in nematics subject to high frustration and 
strong anchoring under conditions of severe confinement.
In our results, planar and homeotropic anchoring conditions behave similarly except for trivial but important differences.
Although the calculation of a complete phase diagram including cavity radius, density 
and particle aspect ratio is beyond our present capabilities, we give general trends as to how the equilibrium
phase depends on these parameters. 

Our theoretical work is mainly inspired by recent experiments on vibrated quasi-monolayers made of granular rods that
interact through approximately overlap forces (hard interactions). Nematic ordering is
observed in these fluids \cite{indios,indios1,Galanis,Aranson,review_hungaro}. Granular materials are non-thermal fluids 
and therefore do not follow equilibrium statistical mechanics. In particular, they flow and diffuse anomalously 
\cite{Aranson1,Yadav}.
However, they can also form steady-state textures that resemble liquid-crystalline states.
In this context, it would be interesting to check whether MC simulation on hard particle models can be useful
to obtain basic trends as to type of patterns, dependence with packing fraction and size of confining cavity, etc. 
The arrangement of rods in a 2D confining cavity has also been investigated in connection with the modelling of actin
filaments in the cell cytoplasm \cite{Mulder}. Self-organised patterns of these filaments have been observed in
various quasi-2D geometries and result from the combined packing and geometrical constraints. Simulation
studies such as the present one could also provide mechanisms to explain this and other experiments on confined 
quasi-two-dimensional nematics \cite{Mottram}. 
 
In Section \ref{SM} we define the particle model, the simulation method and 
provide some details on the analysis. Results are presented in Section \ref{Results},
and a short discussion and the conclusions are given in Section \ref{conclusions}.

\section{Model and simulation method}
\label{SM}

The particle model we use is the hard-rectangle (HR) model, consisting of particles of length-to-width ratio 
$L/\sigma=16$ or $40$ 
that interact through overlap interactions. The configuration of a particle is defined by $({\bm r},\hat{\bm\omega})$,
respectively the position vector of the centre of the particle and the unit vector giving the orientation of the long
particle axis.
A collection of $N$ such particles is placed in a circular cavity of radius $R$. 
We define the packing fraction $\phi$ of the system as the ratio of area covered by rectangles and total area $A$
of the cavity. Thus $\phi=NL\sigma/A=\rho_0L\sigma$, where $\rho_0=N/A$ is the mean density.

For such large length-to-width ratios a fluid of HR undergoes
a phase transition from an I phase to a nematic (N) at rather low densities 
\cite{SCHLACKEN,raton064903,PhysRevE.76.031704,PhysRevE.79.011711,raton014501,PhysRevE.80.011707}. The bulk transition is 
continuous and probably of the Kosterlitz-Thouless type; this detail is irrelevant here since, due to the completely
confined geometry, there can be no true phase transition in the circular cavity and one expects a possibly abrupt but
in any case gradual change from the I phase to the N phase.

The effect of the cavity wall on the particles is represented via an external potential 
$v_{\hbox{\tiny ext}}({\bm r},\hat{\bm\omega})$. In all cases this is a hard potential but, depending on the
type of surface anchoring condition wished (either homeotropic or planar), the potential can be chosen to act on 
the particle centres of mass or on the whole particle --all four
corners of the particle. Specifically,
\begin{eqnarray}
\beta v_{\hbox{\tiny ext}}({\bm r},\hat{\bm\omega})=\left\{\begin{array}{cl}\infty,&\left\{\begin{array}{l}
\cdot\hspace{0.1cm}\hbox{\small at least one corner outside} \\ \hbox{\small cavity (planar)}\\
\cdot\hspace{0.1cm}\hbox{\small centre of mass outside }\\ \hbox{\small cavity (homeotropic)}\end{array}\right\}\\\\                                                                                     
0,&\hbox{otherwise.}\end{array}\right.
\end{eqnarray}
where $\beta=1/kT$, with $k$ Boltzmann's constant and $T$ the temperature. A hard wall acting on the whole particle
promotes planar ordering. However, if the condition is on the particle centers of mass, it is homeotropic anchoring that
is promoted. This was shown by MC \cite{Lowen} and density-functional studies \cite{PhysRevE.79.061703}, and has also 
been confirmed in fluids of hard discorectangles confined in 2D circular cavities or in fluids
of rods confined in slit pores in 3D \cite{doi:10.1080/00268979909483083,heras:4949,0953-8984-19-32-326103}.
Since a single particle close to a wall may have any orientation in this type of condition, one infers that homeotropic 
anchoring results from the collective effect of all particles. By contrast, a hard wall over the whole particle induces 
planar anchoring. In this case anchoring is not the result of a collective effect since a single particle sufficiently 
close to the wall is forced to adopt an orientation parallel to the wall.

The simulation method was the following. We started at low density with a few particles inside the cavity. 
After equilibrating the system using the standard Metropolis algorithm on particle
positions and orientations, a few particles are added and the fluid is equilibrated again. The number of particles added varied between 
$1$ and $20$, resulting in an increase in packing fraction of $\sim 1-10\cdot10^{-3}$ (depending on aspect ratio and cavity 
radius). This process was repeated until a high density was reached.

For the insertion process we first chose one particle at random and create a replica
with the same orientation but with the long axis displaced by $\sim D$. Then we performed a few thousands rotations and 
displacements on the new particle. The addition of one particle, especially when the density is high, may lead to overlap; 
in that case we chose another particle to create the replica and a new attempt was made until insertion was completed 
successfully. The simulation ended when the desired density was reached or if the addition of new particles is no longer possible. 
As usual, a Monte Carlo step (MCS) is defined as an attempt to individually move and rotate all particles in the system.
We performed $5-15\cdot 10^5$ MCS for each $N$. The acceptance probability was set to about $0.2$, 
and depended on the maximum displacement $\Delta r_{\hbox{\tiny max}}$ and maximum rotation $\Delta\phi_{\hbox{\tiny max}}$ each particle is allowed 
to perform in one MCS. Both, $\Delta r_{\hbox{\tiny max}}$ and $\Delta\phi_{\hbox{\tiny max}}$, are adjusted to obtain the desired acceptance 
probability every time we increase the number of particles.

To characterise the fluid structure in the cavity, three local fields are defined: 
(i) A local density $\rho({\bm r})$ in terms of a local packing fraction $\phi({\bm r})=\rho({\bm r})L\sigma$; here
 ${\bm r}=(x,y)$ is the position vector of a particle centre of mass.
(ii) A local order tensor $Q_{ij}({\bm r})$, defined as $Q_{ij}=\left<2\hat{\bm \omega}_i\hat{\bm \omega}_j-\delta_{ij}\right>$,
where $\left<\cdots\right>$ denotes a canonical average, $\hat{\bm \omega}=(\cos{\varphi},\sin{\varphi})$ 
is the unit vector pointing along the particle
axis, and $\varphi$ is the angle with respect to the $x$ axis. The order tensor can be diagonalised, and the largest
eigenvalue $Q({\bm r})$ is taken as the local order parameter (the other eigenvalue is negative and with
the same absolute value). The $x$ axis of the frame where $Q_{ij}$ is diagonal defines a local tilt angle $\psi({\bm r})$ with respect to
the laboratory-frame $x$ axis.

All local quantities, $\phi({\bm r})$, $Q({\bm r})$ and $\psi({\bm r})$ were defined, at each ${\bm r}$, as an average
of each quantity for all particles over a circle of radius $r=0.5L$ and for $2500$ different configurations separated by $50$ 
MCS (all local fields shown in this paper were obtained in this way). This number of configurations is necessary in order to
obtain spatially smooth fields. However, a complication may arise due to the collective rotation of the particles inside the 
cavity. As the cavity does not impose a global director, microstates with the same average order 
parameter but distinct global directors are equivalent. Hence, an ensemble average over these states can artificially 
result in a state with lower order parameter. In particular, the collective rotations of the particles take place more 
frequently at low packing fractions, hindering the nematization analysis. Since the precise location of the transition to 
the nematic state is not the goal of our work, we did not attempt to address this problem.

As shown in section \ref{Results}, distinct states characterised by different local nematic fields
will arise in the cavity as the density is increased. Given the completely restricted geometry and the reduced number 
of particles in the cavity, abrupt changes are not expected and only a continuous transition between the different
states can occur in the system. In order to check this, we have run simulations for selected cavity radii by first 
increasing and then decreasing the number of particles, looking for possible hysteresis effects. 
As expected, no such effects were found in the process and we can be confident that the states described in the 
following section are the stable ones. 

\section{Results}
\label{Results}

An overall picture of the phenomena occurring in the cavity as the density is
increased can be obtained by looking at typical particle configurations, local packing
fraction and local order parameter.
In this section we first present the results for the planar case, and then for the homeotropic case. In both cases
cavity radii from $R=1.0L$ to $10L$ were explored, all for $L/D=16$ and $40$.

\begin{figure*}
\includegraphics[width=\textwidth]{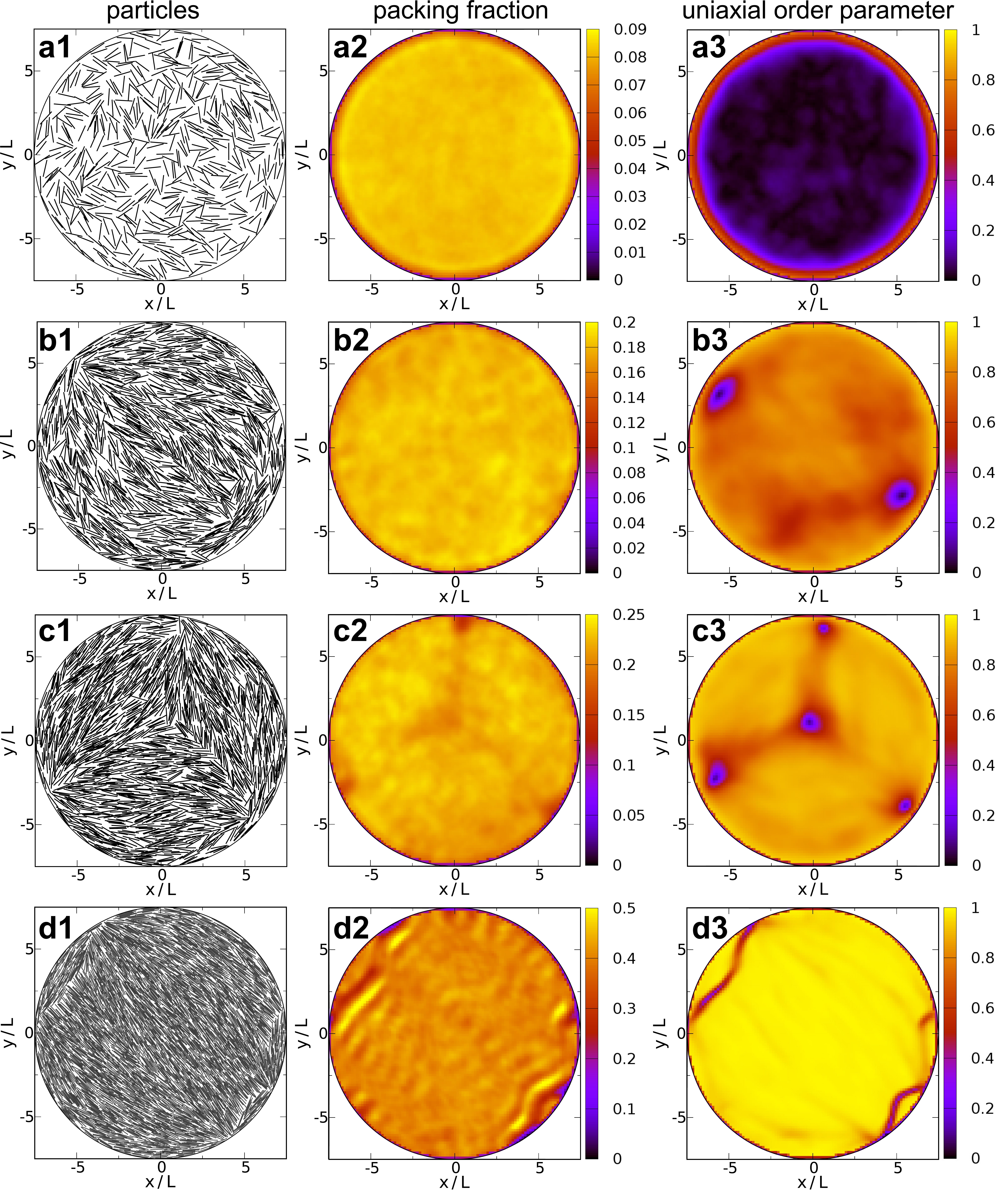}
\caption{Cavity with radius $R=7.5L$, planar anchoring conditions and particle aspect ratio $L/D=40$. 
Left column: snapshot of particle configurations. Middle column: local packing fraction. Right column: 
local order parameter. First row: I phase, with $N=570$ and global packing fraction 
$\phi\simeq 0.08$. Second row: N$_2$ phase, with $N=1270$ and $\phi\simeq 0.18$.
Third row: a probably metastable phase, with $N=1550$ and $\phi\simeq 0.22$. Forth row: N$_u$ phase, with 
$N=2820$ and $\phi\simeq 0.40$.}
\label{fig2}
\end{figure*}

\subsection{Planar anchoring}  

Representative results in the case of planar anchoring can be obtained from the case $R=7.5L$ and 
$L/D=40$ (the behaviour is qualitatively the same for the other cavity radii and aspect ratios analysed). 
The initial number of particles was $N=250$ (corresponding to a packing fraction $\phi\simeq 0.04$), 
and the final number of particles was $N=3000$ ($\phi\simeq 0.42$). The results are shown in Fig. \ref{fig2}, where each row corresponds to a given packing fraction,
increasing from top to bottom.

At low $\phi$ (first row in Fig. \ref{fig2}) the fluid is disordered (I phase), except at a thin film next to the wall 
which presents some degree of planar ordering. As the fluid becomes more dense it 
undergoes a quasi-transition from the I phase to the N$_2$ phase at a density close to the bulk transition (the 
second row in the figure, for $\phi=0.180$, corresponds to a nematic state, i.e. beyond the bulk transition).
This density agrees closely with that predicted for hard rods in 2D \cite{Bates,raton064903}. Nematization in
the case $L/D=16$ is qualitatively similar, except that the transition density is more or less doubled.

Once a nematic fluid is established in the cavity, the local director is subject to frustration due to
the geometry. The planar surface 
orientation is satisfied by the particles but, due to the topological restrictions imposed by the wall, the nematic fluid 
creates two disclinations of topological charge $+1/2$ next to the walls in diametrically opposed regions. 
This feature manifests itself in panel b3 through the depleted order parameter at the disclination cores
(by contrast, the local $\phi$, panel b2, is not sensitive because the ordered and disordered phases have the same 
density). Two isolated $+1/2$ disclinations are always more stable than a single point defect of charge $+1$ because the 
free energy is proportional to the square of the topological charge. The boundaries could modify this balance, but our 
results, already predicted by density-functional calculations \cite{PhysRevE.79.061703}, indicate that this is not 
the case.

Sometimes along the MC chain, configurations with
two extra disclinations are excited (third row in Fig. \ref{fig2}): one of charge $+1/2$, close to the surface and forming 
an equilateral triangle with the previous two, and another one with charge $-1/2$ at the centre 
(panel c3); these configurations, which still have a total topological charge of $+1$, do not appear very often in the 
MC chain since they involve a higher elastic free energy, and anyway the $-1/2$ and $+1/2$ disclinations tend to 
annihilate each other. A similar metastable configuration has been observed in simulations of hard spherocylinders lying on the 
surface of a three-dimensional sphere \cite{Lowen}.

At higher densities a dramatic structural change can be observed (fourth row in Fig. \ref{fig2}). As $\phi$ increases,
elastic stresses becomes very large because of the strong dependence of elastic constants, $K$, with density.
As a consequence, a quasi-uniform director configuration (N$_u$ phase) with little elastic stress is formed
beyond some critical value $\phi_c$. 
The director orientation is not completely uniform. A perfectly uniform director configuration would imply that the 
planar orientation favoured at the wall is not completely satisfied. However, the fluid can reduce the increased 
surface free-energy implied by a strictly uniform director field by creating two fluctuating domain walls, panel d3,
that define two diametrically opposed domains where the director rotates by $90^{\circ}$. Alternatively, we can view this 
structure as a polar structure (stable at lower densities) where the two point defects are smeared out into a curved 
one-dimensional interface. Particles in the two small domains satisfy the surface orientation. Note that the domain walls 
behave as a soft wall: particles of the central domain next to the interface are highly ordered and the density in these 
regions is increased (panel d2). Domain walls are seen to behave as highly fluctuating structures.\\

\begin{figure}
\includegraphics[width=1.0\columnwidth]{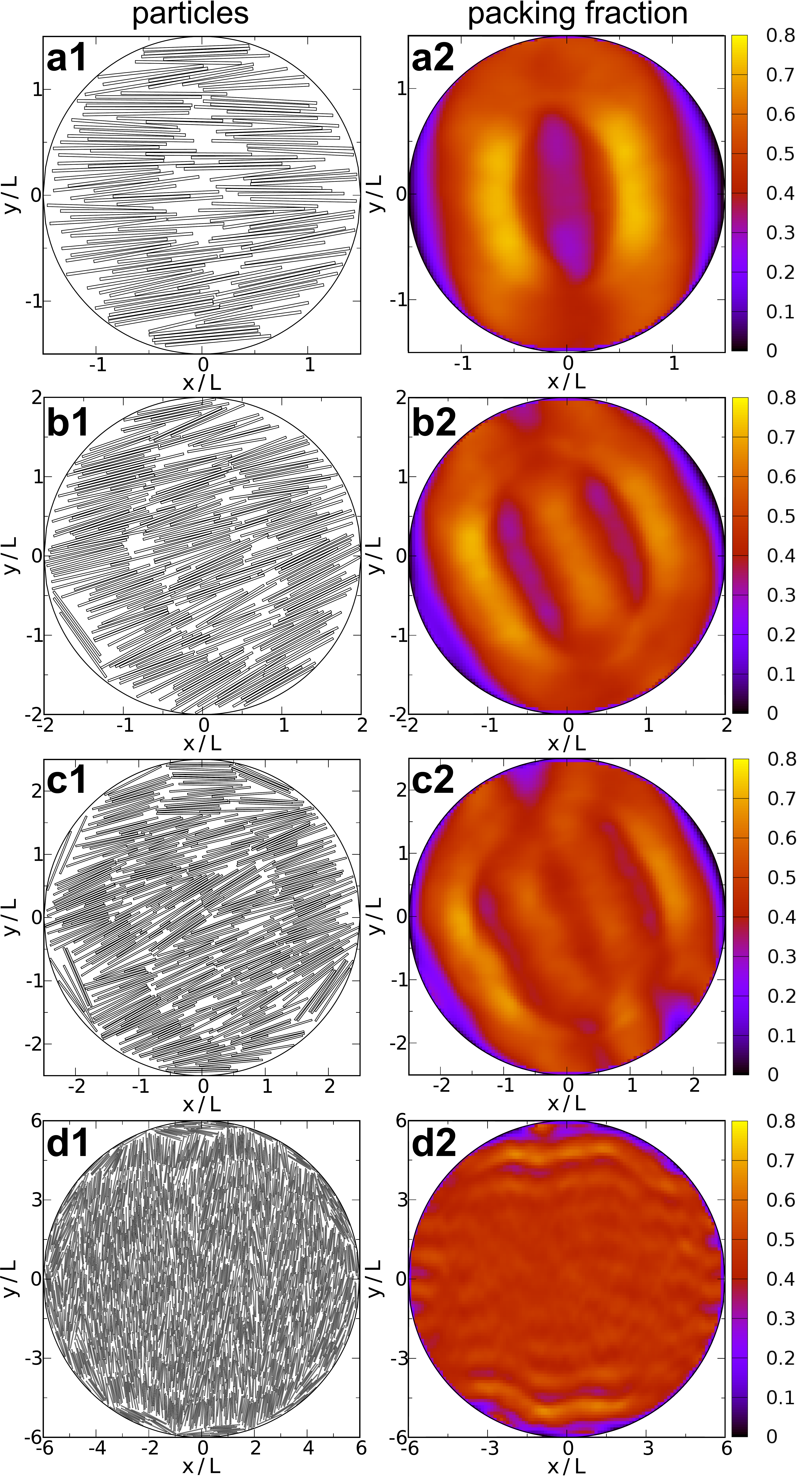}
\caption{High-density states of particles with $L/D=40$ confined in cavities with planar anchoring and 
different radii. The global packing fraction is $\phi\simeq 0.45$. Left column: representative snapshots of particle 
configurations. Right column: local packing fraction. Cavity radius increases from top to bottom. 
First row: $R=1.5L$, $N=126$. Second row: $R=2.0L$, $N=224$. Third row: $R=2.5L$, $N=350$. 
Fourth row: $R=6.0L$, $N=2025$.}
\label{fig3}
\end{figure}

The value of packing fraction, $\phi_c$, at which the N$_2$-N$_u$ transition takes place
increases with cavity radius $R$. To
understand this, let us consider the free energy $F$ of both configurations, N$_2$ and N$_u$, with respect to an 
undistorted nematic state with free energy $F_0$. Three terms contribute to the excess
free-energy $\Delta F=F-F_0$: domain walls, $F_w$, elastic deformations of the 
director field, $F_e$, and disclination cores, $F_c$. In the N$_2$ state the director field is distorted; the free energy 
presents a logarithmic dependence with $R$, i.e. $F_e\sim K(\phi)\log{R}$ \cite{dephysics,kleman2002soft}, 
while the contribution from the two disclinations is almost constant (assuming the distance between the cores to be
independent of $R$). In the N$_u$ state director deformations 
are negligible in comparison to the other phase, but the presence of domain walls increases the free energy. 
Since $F_w$ is proportional to the length of the domain wall, it should also increase with $R$, but faster than $F_e$. 
Due to the weaker dependence of $F_e$ with $R$, we expect a transition from N$_u$ to N$_2$ 
when $R$ is increased beyond some critical value and consequently $\phi_c(R)$ should be an increasing function of $R$ in view
of the density dependence of the elastic constants $K(\phi)$. This conclusion is confirmed by our simulations (not shown).

\begin{figure*}
\includegraphics[width=\textwidth]{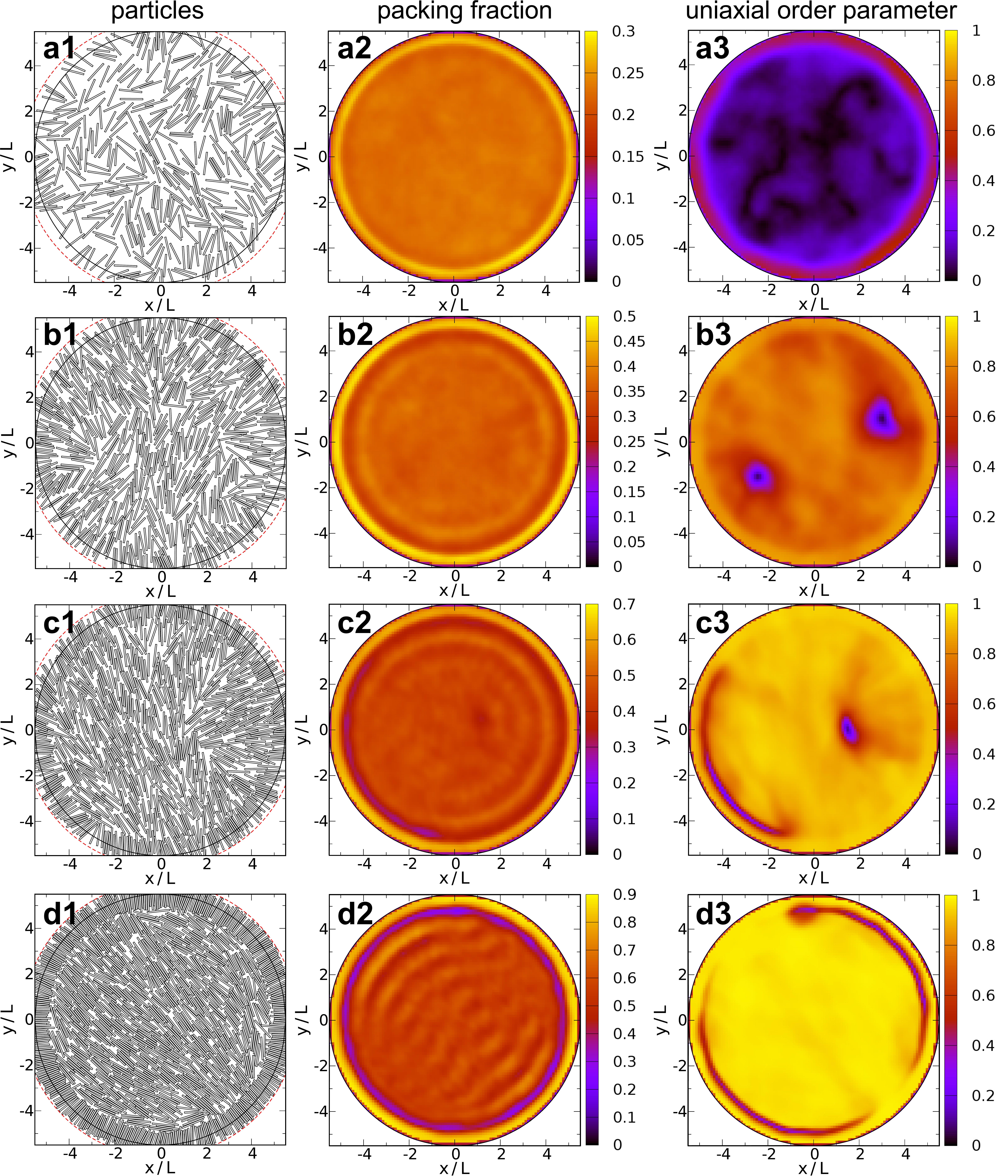}
\caption{Cavity with radius $R=5.5L$, homeotropic anchoring conditions and particle aspect ratio 
$L/D=16$. Left column: snapshot of particles configurations (circle in black is the actual cavity wall of radius $R$, 
while circle in red represents the effective cavity with radius $R_{\hbox{\tiny eff}}$). Middle column: 
local packing fraction. Right column: local order parameter. First row: I phase with $N=378$ and global effective packing 
fraction $\phi_{\hbox{\tiny eff}}\simeq 0.21$. Second row: N$_2$ phase with $N=618$ and
 $\phi_{\hbox{\tiny eff}}\simeq 0.34$.
Third row: a state intermediate between the N$_2$ and N$_u$ phases, with $N=778$ and $\phi_{\hbox{\tiny eff}}\simeq 0.43$. 
Forth row: N$_u$ phase with $N=1022$ and $\phi_{\hbox{\tiny eff}}\simeq 0.56$.}
\label{fig4}
\end{figure*}

\begin{figure*}
\includegraphics[width=\textwidth]{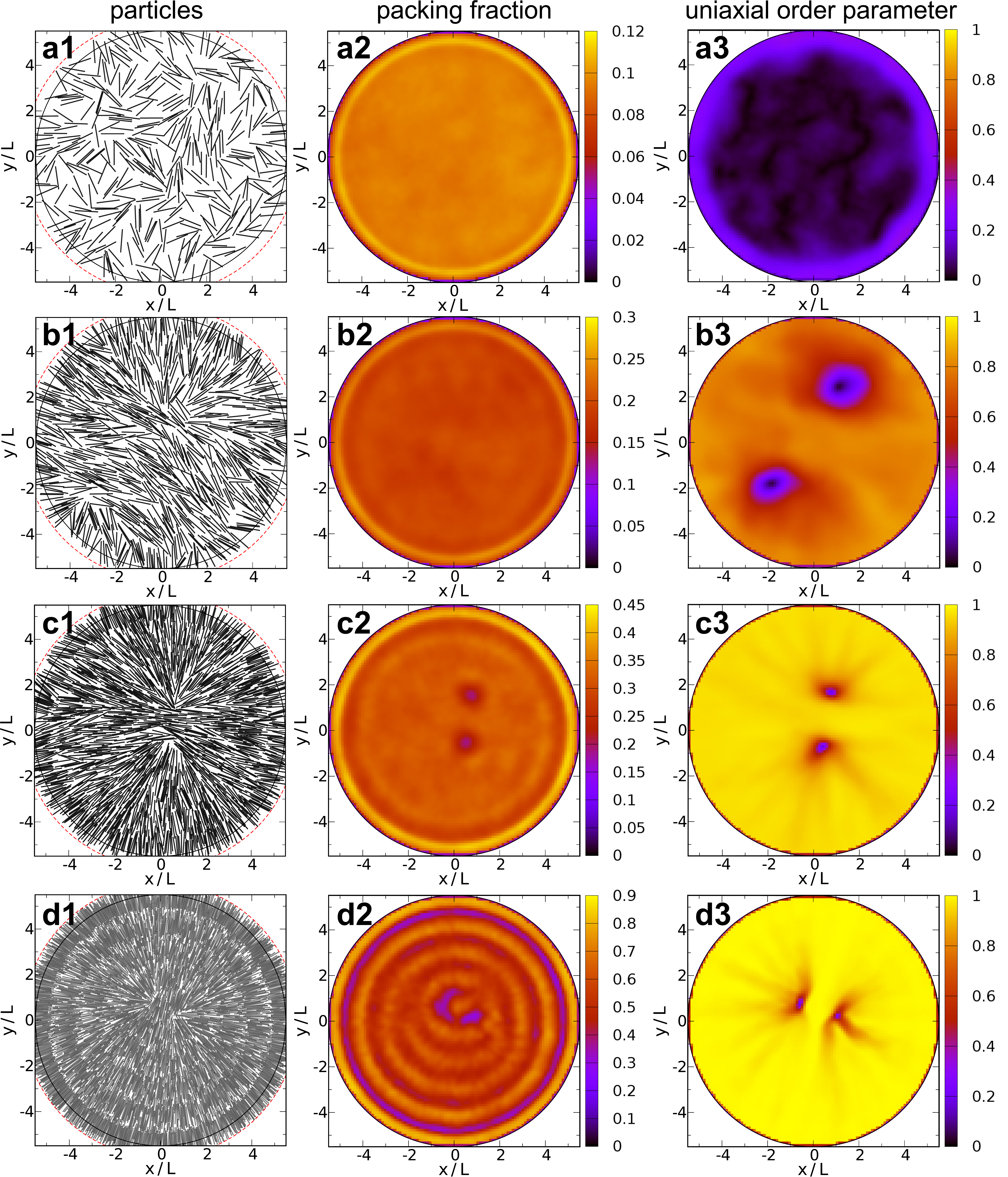}
\caption{Cavity with radius $R=5.5L$, homeotropic anchoring conditions and particle aspect ratio 
$L/D=40$. Left column: snapshot of the particle configurations (circle in black is the actual cavity wall of radius $R$, 
while circle in red represents the effective cavity with radius $R_{\hbox{\tiny eff}}$). 
Middle column: local packing fraction. Right column: 
local order parameter. First row: I phase with $N=385$ and average effective packing fraction 
$\phi_{\hbox{\tiny eff}}\simeq 0.09$. Second row: N$_2$ phase with $N=805$ and $\phi_{\hbox{\tiny eff}}\simeq 0.18$.
Third row: N$_2$ phase with $N=1325$ and $\phi_{\hbox{\tiny eff}}\simeq 0.29$. Forth row: phase exhibiting layering
in the radial direction, with $N=2445$ and $\phi_{\hbox{\tiny eff}}\simeq 0.54$.}
\label{fig5}
\end{figure*}

\begin{figure*}
\includegraphics[width=6.2in]{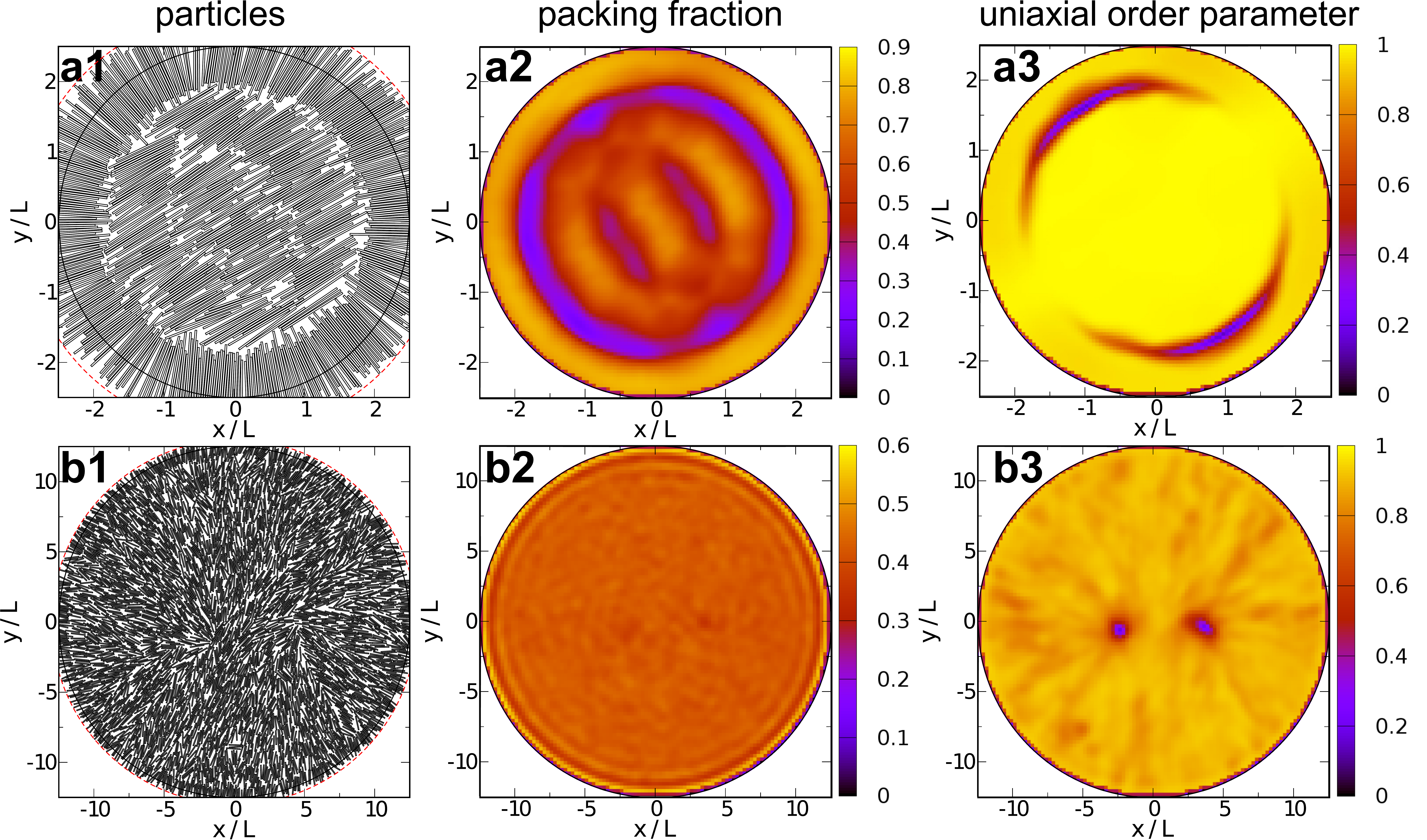}
\caption{High density states in a cavity with homeotropic anchoring. Left column: 
representative particle configurations. Middle column: local packing fraction. Right column: 
local order parameter. First row: uniform phase, 
$L/D=40$, $R=2.5L$, $N=610$, $\phi_{\hbox{\tiny eff}}\simeq 0.54$. Second row: polar phase, 
$L/D=16$, $R=12.5L$, $N=3508$, $\phi_{\hbox{\tiny eff}}\simeq 0.41$.}
\label{fig6}
\end{figure*}

On further increasing the value of $\phi$, the fluid may develop smectic-like 
layers, reflecting the corresponding transition in bulk. The role played by cavity radius is especially important in 
this regime. To see this, we plot in Fig. \ref{fig3} representative snapshots of the particle configurations (left column) and the local packing 
fraction (right column) of particles with aspect ratio $L/D=40$.
In all cases the average packing fraction is $\phi\simeq 0.45$ (well above the bulk I-N transition) and the radius 
increases from top to bottom: $R=1.5L$, $2L$, $2.5L$ and $6L$. 
As expected, strong commensuration effects arise in the cavity at high density. For small cavity radii, the particles 
form well defined layers (first three rows in figure), the number of which depends on the available space. The formation 
of layers inside the cavity is the analogue of capillary smectization of a liquid crystal in slab geometry previously 
analysed in 3D \cite{PhysRevLett.94.017801,PhysRevE.74.011709,0953-8984-20-42-425221,0953-8984-22-17-175002} and
2D \cite{Yuri}. In general, the circular shape of the cavity frustrates the formation of well-defined layers,
but the combined effect of shape and size may frustrate or enhance the formation of layers. In this system
small cavity sizes promote layering: in the cases shown, the fluid remains 
in a nematic-like N$_{\hbox{\small u}}$ state when $R=6L$ (see panels d1 and d2 of Fig. \ref{fig3}), but
for smaller cavities at the same packing fraction well defined smectic-like layers can develop.

\subsection{Homeotropic anchoring}

In this case the wall acts as a hard wall on the particle centres of mass. In order to be able to compare with the
planar case, we define an effective cavity radius, $R_{\hbox{\tiny eff}}=R+\sqrt{L^2+D^2}/2$,
and obtain an effective packing fraction as $\phi_{\hbox{\tiny eff}}=NL\sigma/(\pi R_{\hbox{\tiny eff}}^2)$. 
In contrast with the planar case, for homeotropic anchoring we observe strong differences with respect to particle aspect 
ratio for a fixed cavity radius: the N$_u$ phase is stabilised for $L/D=16$, but the N$_2$--N$_u$ transition
is preempted by the formation of smectic-like layers when $L/D=40$ and, as a result, no quasi-uniform
configuration occurs.

Fig. \ref{fig4} summarises a typical evolution of the configurations as $\phi$ is increased when $L/D=16$.
The low-density configuration (first row)
is similar to the planar case: a thin (one-particle thick) film develops
at the wall, now with normal average orientation of the particles, while the rest is disordered. 

When the density increases and nematic order appears in the whole cavity (second row of Fig. \ref{fig4}), 
the topological constraints force the creation of two disclinations 
of charge $+1/2$; this is as in the planar case (second row of Fig. \ref{fig2}).
Here the two defects are not at the wall but a bit separated. This feature was predicted by 
density-functional theory \cite{PhysRevE.79.061703,herasliqcrys} and results from the effective repulsion of the 
defect by the wall combined with the mutual repulsion between the two defects (in the planar case such defect-wall
repulsion does not exist). As the density is increased further
the two disclinations can be seen to approach each other (not shown). 
This behaviour is in contrast with that predicted by the Onsager-like theory 
analysed in \cite{PhysRevE.79.061703}, according to which the relative distance should increase with
chemical potential (or, equivalently, density). Again two effects are at work as density increases: 
defect-wall repulsion, which increases due to the strong spatial ordering near the wall and the incipient stratification
from the wall in the normal direction (panels b2), and defect mutual repulsion, which also increases due to
the higher elastic stiffness of the nematic. The poor description of strong density modulations occurring at high
density in Onsager theory may explain the discrepancy.

The third row in Fig. \ref{fig4}, corresponding to a larger value of packing fraction, represents an intermediate 
(probably metastable) stage between the polar state and what appears to be the stable configuration at high density:
the uniform phase, N$_u$ (fourth row). Here the director alignment in the cavity is more or less uniform, with little
bend-like elastic distortion. However particles in the first layer, highly packed in a compact normal configuration, 
form a well defined and stable film acting as a soft wall favouring planar orientation. This results in a 
quasi-uniform phase N$_u$, similar to that found in the case of planar anchoring at high densities. The resulting 
configuration contains two extended domain walls separating the first layer from the central nematic domain. 
The N$_2\to$N$_u$ transition is driven by an anchoring-transition mechanism: the orientation of particles 
next to the first layer changes from normal (see panel b1 in Fig.\ref{fig4}) to tangential (panel d1) along opposite
arcs spanning $\sim 120^{\circ}$; note that in the configuration of panel c1 the transition has taken place only in one side.       
As in the N$_2$ phase, the formation of these domain walls breaks rotational symmetry and establishes a direction
along which smectic-like layers can grow at higher densities (see incipient layering in panel d2). 

The stability of the N$_u$ phase was checked by preparing a configuration of $N=900$ rods in a 
cavity of radius $R=5.5L$ containing a central radial disclination of charge $+1$. During the MC chain the 
system rapidly formed a polar phase with two $+1/2$ disclinations, similar to that depicted in Fig. \ref{fig4} (b1).
Then an anchoring transition took place in half of the cavity, and finally the N$_u$ phase was stabilised and
remained unchanged for the rest of the simulation. This can be taken as strong evidence that the N$_u$ phase 
is the truly stable phase. 

For particle aspect ratio $L/D=40$ and the same cavity radius, Fig. \ref{fig5}, low-density states are similar to those
with aspect ratio $L/D=16$ (first and second row of Fig. \ref{fig4}). However,
the order parameter close to the surface is significantly lower than in the case of planar 
anchoring (compare panels a3 of Figs. \ref{fig2} and \ref{fig5}). The reason is the following:
A single particle at close contact with the wall has a specific (tangential) alignment when the wall acts
over the whole particle; out-of-tangential alignments lead to particle-wall overlap and are not allowed,
which increases the order parameter. In contrast, a hard wall acting on the centers of mass does not induce any 
favoured orientation on the particles if the density is sufficiently low.

As $\phi$ increases a nematization transition occurs, and a N$_2$ phase is stabilised at intermediate densities. However, for larger densities, the situation changes dramatically:
the homeotropic anchoring imposed by the surfaces always propagates to the inner cavity since the locking mechanism
that anchors particles to the first layer is much more effective: the normal-to-tangential anchoring transition never 
takes place. At higher packing fractions, $\phi_{\hbox{\tiny eff}}\simeq 0.54$ (fourth row in Fig. \ref{fig4}), a density
stratification grows from the surface and particles form well-defined, concentric layers, the two $+1/2$ disclinations
being pushed to the central region. The size of the defect cores (regions where the order parameter vanishes) 
becomes significantly smaller as density is increased. Therefore, the absence of the anchoring transition inhibits
the formation of the N$_u$ phase and the fluid instead goes directly to a smectic-like phase exhibiting concentric layers.

Thus, for $R=5.5L$, the high-density states are completely different depending on the aspect ratio.
Anchoring is much stronger in the case $L/D=40$, the anchoring transition does not take place, 
and the quasi-uniform configuration is inhibited as a result. At high densities, smectic-like layers do not grow along
a fixed direction but in the radial direction instead. 

From this evidence, it may seem that the value of aspect ratio $L/D$ determines 
the type of structure in the cavity as $\phi$ increases. This is not the case, and in fact the $R/L$ ratio is a more 
important factor. To check this, we have conducted simulations for both elongations but with very different cavity sizes,
$R=2.5L$ for $L/D=40$ and $R=12.5L$ for $L/D=16$, Fig. \ref{fig6}. In the first case a uniform configuration with 
domain walls is stabilised; in the second, no anchoring transition occurs and, as a consequence, the N$_u$ phase is not
stabilised before layering sets in. We conclude that a large value of $R/L$ favours anchoring and inhibits the anchoring
transition and the formation of the quasi-uniform phase. In turn, the critical value of the ratio $R/L$ separating both
regimes depends on $L/D$ to some extent.

\section{Discussion and conclusions}
\label{conclusions}

In summary, we have observed the formation of domain walls in 2D nematic fluids subject
to frustration as a result of confinement in small cavities. These structures, not predicted by elasticity theory, 
are similar to the step-like defects obtained in model 3D nematics confined in hybrid planar slit pores. 
At domain walls the director orientation changes at the molecular scale, so that the elastic field becomes singular 
along extended interfaces (lines in 2D or surfaces in 3D). 
In this way the elastic distortion is greatly relaxed, while the surface orientation is still optimised. Domain walls
were predicted in an analysis of the neighbourhood of half-integer nematic disclinations, using Landau-de Gennes theory 
\cite{Schopohl}. In this case the domain wall is a way to avoid a disordered defect core. In hybrid planar slit pores, 
where the two facing surfaces favour antagonistic directions, a domain-wall structure has also been predicted by 
Landau-de Gennes theory \cite{PM} and by density-functional calculations \cite{PhysRevE.79.011712}.
 
Here we have extended the observation of these domain walls to real particle simulations of confined nematics in 2D. 
In our system and for a fixed cavity radius, we have found the sequence I$\to$N$_2\to$N$_u$ for increasing packing fraction 
$\phi$. The packing fraction at which the N$_2\to$N$_u$ transition occurs strongly depends on cavity radius $R$. At high 
density smectic-like layers are formed in the fluid. The N$_u$ phase is a quasi-uniform phase with little director distortion; this
is realised by the creation of two domain walls which help maintain the favoured surface orientation in the whole cavity.
Homeotropic anchoring conditions induce the formation of a highly-packed surface film, which may drive an anchoring transition 
to a tangential orientation and the formation of a large quasi-uniform nematic domain separated from the first layer
by domain walls. Large $R/L$ ratios inhibit the anchoring transition and therefore the stabilisation of the N$_u$ phase,
but the critical value of $R/L$ depends on the aspect ratio $L/D$. Our results emphasise the possibility of domain-wall
formation in small confined systems as a mechanism to optimise surface and elastic stresses. 
 
It is interesting to compare our results with other studies. Dzubiella et al. \cite{Lowen} analysed a similar system
using MC simulation. They focused on moderate densities and obtained the N$_2$ phase. Galanis et al. \cite{Galanis} studied
 vibrated quasi-monolayers of rods in circular and square geometries, and observed the formation of nematic patterns. 
The patterns were seen to be well described by continuum elastic 
theory, but the particle configurations of the experiment seem to exhibit some evidence of domain walls not contemplated
in that work. The formation of two $+1/2$ disclinations has been also predicted in active matter confined in cylindrical capillaries 
\cite{PhysRevLett.110.026001} and circular cavities \cite{PhysRevLett.109.168105}.

Recently, liquid-crystalline ordering has been studied in square cavities using density-functional 
\cite{Miguel} and MC simulation \cite{Dani}. Domain walls can be stabilised in these systems. For some geometries the
formation of domain walls may be a necessary requirement for the development of confined phases with spatial order 
(smectic or columnar) at higher densities. 

\section{Acknowledgments}
We thank M. Schmidt, T. Geigenfeind, S. Rosenzweig and Y. Mart\'{\i}nez-Rat\'on for useful discussions and comments. E. V.
 acknowledges financial support from Programme MODELICO-CM/S2009ESP-1691 (Comunidad Aut\'onoma de Madrid, Spain), and FIS2010-22047-C01  (MINECO, Spain).

\end{document}